\documentstyle[11pt,fleqn,epsf]{article}
\def\ap#1#2#3{     {\it Ann. Phys. (NY) }{\bf{#1},} (#3) #2 }
\def\jmp#1#2#3{   {\it J. Math. Phys. } {\bf{#1},}  (#3) #2 }

\def\jpa#1#2#3{   {\it J. Phys. A: Math. Gen.} { \bf{#1},} (#3) #2 }

\def\pla#1#2#3{    {\it Phys. Lett. }{ A \bf{#1},} (#3) #2 }
\def\npb#1#2#3{    {\it Nucl. Phys. }{ B \bf{#1},}  (#3) #2 }

\def\mpla#1#2#3{    {\it Mod. Phys. Lett. }{A \bf{#1},} (#3) #2 }

\def\cjp#1#2#3{    {\it Can. J. Phys. }{\bf{#1},} (#3) #2 }
\def\eq#1{{equation~(\ref{#1})}}

\newcommand{\bea}{\begin{eqnarray}}
\newcommand{\beq}{\begin{equation}}
\newcommand{\eea}{\end{eqnarray}}
\newcommand{\eeq}{\end{equation}}
\newcommand{\nnu}{\nonumber}

\begin{document}
%%%%%%%%%%%%%%%%%%%%%%%%%
\title{\bf High-Precision Numerical Determination of Eigenvalues for
a Double-Well Potential Related to the Zinn-Justin  Conjecture}
{\small
\author{H. A. Alhendi$^1$\ and \ E. I. Lashin$^{1,2}$\\
$^1$ Department of physics and Astronomy, College of Science,\\ King Saud University, Riyadh,
Saudi Arabia \\
$^2$ Department of Physics, Faculty of Science, \\Ain Shams University, Cairo, Egypt\\
Email: alhendi@ksu.edu.sa, lashin@ksu.edu.sa}
}
\maketitle
\begin{abstract}
A numerical method of high precision is used to calculate the
energy eigenvalues and eigenfunctions for a symmetric double-well
potential. The method is based on enclosing the system within two
infinite walls with a large but finite separation and developing a
power series solution for the Schr$\ddot{\mbox{o}}$dinger
equation. The obtained numerical results are compared with those
obtained on the basis of the Zinn-Justin conjecture and found to
be in an excellent agreement.
\end{abstract}
\section{Introduction}
Quantum mechanical tunneling through finite barriers is a well
established phenomenon in theory and application. The symmetric
double well potential is one of the many examples exhibiting this
phenomenon. In this case, the energy splitting generated by tunneling
can be estimated with the help of the well-known semi-classical
WKB approximation and instanton techniques (see for example
\cite{col}). However, to calculate this splitting accurately,
one needs an effective method of high precision.

In a series of papers, Zinn-Justin \cite{zinn1}
developed a  conjecture (to be termed
'the Zinn-Justin conjecture') to determine the energy levels of a quantum
Hamiltonian H, in cases where the potential has degenerate minima.
This conjecture takes the form of the generalized Bohr-Sommerfeld
quantization formulae. It has been applied, among other potentials, to the
case of the symmetric double well.
In this case the Hamiltonian is
\beq
H=-{g\over 2} {\partial^2\over \partial q^2} + {1\over g}
V(q),\hspace{1cm}\mbox{where}\hspace{1cm} V(q)={1\over 2} q^2
(1-q)^2.
\label{ham}
\eeq
It is obvious that this Hamltonian is invariant under the transformation $(q\rightarrow
1-q)$.
The energy eigenvalues for this potential can be obtained by finding  a solution to the
Zinn-Justin conjecture equation:
\beq
{1\over \sqrt{2\,\pi}} \Gamma\left({1\over 2}- D(E,g)\right)
\left(-{2\over g}\right)^{D(E,g)}\;\exp{\left[-A(E,g)/2\right]}=\pm
i.
\label{forma1}
\eeq
The function $D(E,g)$ has a perturbative expansion in powers of $g$, of which
the first few terms are
\beq
D(E,g)= E + g\left(3 E^2+{1\over 4}\right) + g^2 \left(35 E^3 +
{25\over 4} E \right) + O(g^3).
\label{funa}
\eeq
The other function $A(E,g)$ receives
contributions from the instanton  expansion in the path integral
and its first few terms are
\beq
A(E,g)= {1\over 3 g} + g\left(17 E^2 + {19\over 12}\right) +
g^2\left(227 E^3 + {187\over 4} E\right) + O(g^3).
\label{funb}
\eeq

The energy $E_{N,\pm}$ can be extracted from \eq{forma1} by  expanding
in powers of $g$ and in the two quantities
\beq
\lambda (g) = \ln{\left(-{2\over g}\right)}\;\;\;\; \mbox{and}\;\;\;\; \xi (g) =
{\exp{\left[-{1/(6 g)}\right]}\over \sqrt{\pi\,g}}.
\label{par1}
\eeq

The complete semi-classical expansion of $E_{N,\pm}$ has the form
\cite{zinnrev}
\beq
E_{\pm,N}(g)=\sum_{l=0}^{\infty} E_{N,l}^{(0)} g^l
+\sum_{n=1}^{\infty} \left({2\over g}\right)^{N n}
\left((\mp){e^{-1/{6 g}}\over \sqrt{\pi g}}\right)^n
\sum_{k=0}^{n-1} \left(\ln{(-2/g)}\right)^k
\sum_{l=0}^{\infty}\epsilon_{n k l}^{(N,\pm)} g^l.
\label{enexp}
\eeq
The coefficients $\epsilon$ relevant to the numerical calculation have been
explicitly calculated in \cite{zinnlett}. The number $N$ is the
unperturbed quantum number which corresponds to
\beq
E_{\pm,N}(g)= N+1/2 + O(g).
\label{defN}
\eeq
A detailed  exposition of the above equations can be found in \cite{zinnrev}.

In \cite{zinnlett}, numerical calculations have been carried out
and led to the energy eigenvalues for the ground
and the first excited states respectively, for $g=0.001$,
\bea
E_{0,+}(0.001)&=& 0.49899\;54548\;62109\;17168\;91308\;39481\;92163
                  \;68209\;47240\;20809\nnu\\
              & &   \;\;\;66532\;93278\;69722\;01391\;
              \underline{15135\;28505\;38294\;45798\;45759\;95999}\nnu\\
              & &   \;\;\;
              \underline{06739\;55175\;84722\;67802\;81306\;96906\;01325\;25943\;
              77289\;94365}\nnu\\
              & &   \;\;\;\underline{88255\;24440\;17437\;12789\;27978\;99793},
\eea
\bea
E_{0,-} (0.001)&=& 0.49899\;54548\;62109\;17168\;91308\;39481\;92163\;68209\;
                     47240\;20809\nnu\\
              & &   \;\;\;66532\;93278\;69722\;01391\;
              \underline{29839\;92959\;55803\;70812\;27749\;92448}\nnu\\
              & &   \;\;\;
              \underline{48259\;36743\;64757\;68328\;84835\;35511\;34663\;
                         06309\;82331\;51885}\nnu\\
              & &
              \;\;\;\underline{23308\;08622\;84780\;52722\;10103\;67282}.
\eea
The above numerical results have been obtained by lattice
extrapolation using a modified Richardson algorithm \cite{zinnlett}.

This tiny difference encourages us to seek for
an independent but simple and direct method, which allows us to obtain
the energy eigenvalues for the potential in \eq{ham} and compare them
with the above numerical results.
In addition, the present method allows us to
obtain an accurate description for the corresponding wavefunctions.
This method has been previously
applied to various potential functions with and without
degenerate minima, leading to results with high accuracy \cite{ourdw}.

The method, as will be described in the next two sections, is based on
power series solution of the Schr$\ddot{\mbox{o}}$dinger equation in a
finite range. It has appeared from time to time in the literature
\cite{sec,bar,kil}, but has not been developed to its maximum
efficiency. We shall show that, by using the computer algebra
systems (for example Mathemtica) which can deal with exact
numbers, the accuracy of the method can be substantially improved.

In the following section, for illustrative purpose, we explain our
method by applying it to the well-known exactly solvable harmonic
oscillator potential and then extend it to the symmetric double well.

\section{Calculations and Results}
In this section  we, first, consider the well-known exactly solvable harmonic
oscillator. In this case,
the Schr$\ddot{\mbox{o}}$dinger equation reads $(\hbar =1, m=1)$
\beq
\left[-{1\over 2} \frac{d^2}{d\,q^2}+E-V(q)\right]\Psi(q)=0,
 \label{scheq0}
 \eeq
where
\beq
V(q)={1\over 2}\; q^2.
\eeq
The exact energy eigenvalues and the corresponding eigenfunctions
are
\bea
E_N &=& \left(N+{1\over 2}\right), \hspace{1cm} N = 0,1,2,\cdots , \nnu\\
\Psi_N(q)& = & 2^{-{N\over 2}}\;(N!)^{-{1\over 2}}\;\pi^{-{1\over 4}}\;
\exp{\left(-{q^2\over 2}\right)}\; H_N(q)\nnu , \\
\eea
where $H_N(q)$ are the Hermite polynomials.

For the harmonic oscillator confined between two infinite walls at
$q=\pm\;L$, we develop a power series solution in the form
\beq
\Psi(q)=\sum_{i=0}^{\infty} a_i\, q^{i}.
\label{sol0}
\eeq
Substituting in \eq{scheq0}, one gets the following
recursion relation:
\beq
a_i={a_{i-4} - 2\,E\, a_{i-2}\over i\,(i-1)},
\hspace{3mm} i\neq 0,1 \hspace{3mm}\mbox{and}\;\; a_i = 0
\;\;\; \mbox{when}\;\;\; i<0.
\label{recur0}
\eeq
The symmetry of the potential
implies that we have two types of solutions: the even solutions
obtained by imposing (ignoring normalization) $a_0=1,a_1=0$ and
the odd ones by imposing $a_0=0, a_1=1$. The energy
eigenvalues are then obtained from the condition $\Psi(E,L)=0$ for
both cases.

For numerical calculations, we approximate the power series in
\eq{sol0} with a truncated one having a finite number of terms $\Psi_I(E,q)$,
where $I$ is the number of non-vanishing terms. The boundary condition
for a specific value of $L$ corresponds to $\Psi_I(E,L)=0$. To get the
zeros of $\Psi_I(E,L)$ with respect to $E$, we first plot a graph for $\Psi_I(L,E)$
as a function of $E$ to locate where $\Psi_I(L,E)$ changes sign.
We then can use two nearby  points containing one single root as the initial iteration for
the 'bisection method' to find the zeros. In doing this we have used Mathematica
package version~3 and also have relied extensively on its ability
to manipulate exact numbers. The stability of the numerical results
to a certain degree of accuracy is checked, for a particular $L$, by
increasing $I$ till the obtained value of $E$ stays fixed.

\begin{figure}[hbt]
\centering
\begin{minipage}[c]{0.5\textwidth}
\epsfxsize=7cm
\centerline{\epsfbox{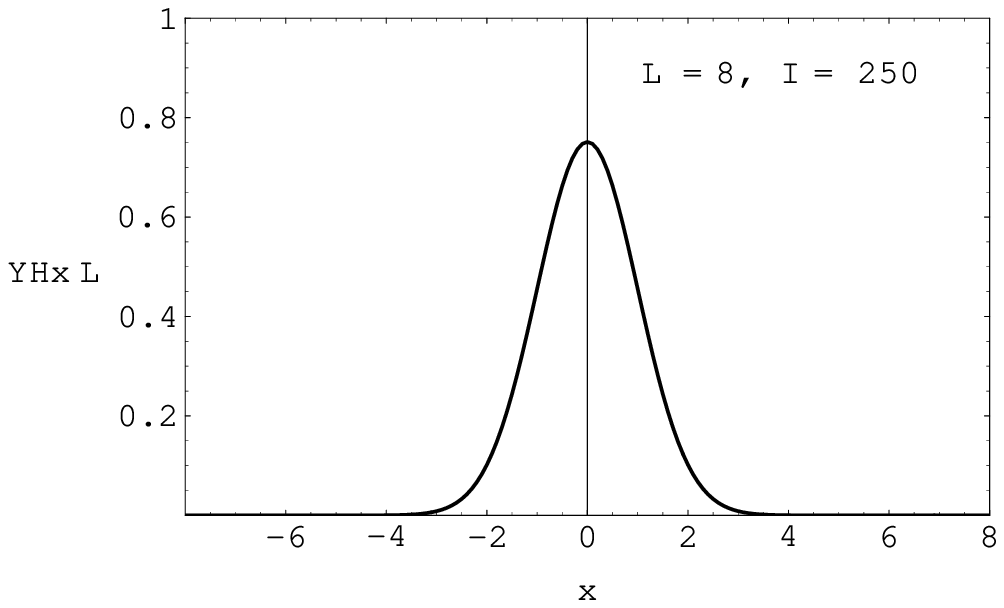}}
\end{minipage}%
\begin{minipage}[c]{0.5\textwidth}
\epsfxsize=7cm
\centerline{\epsfbox{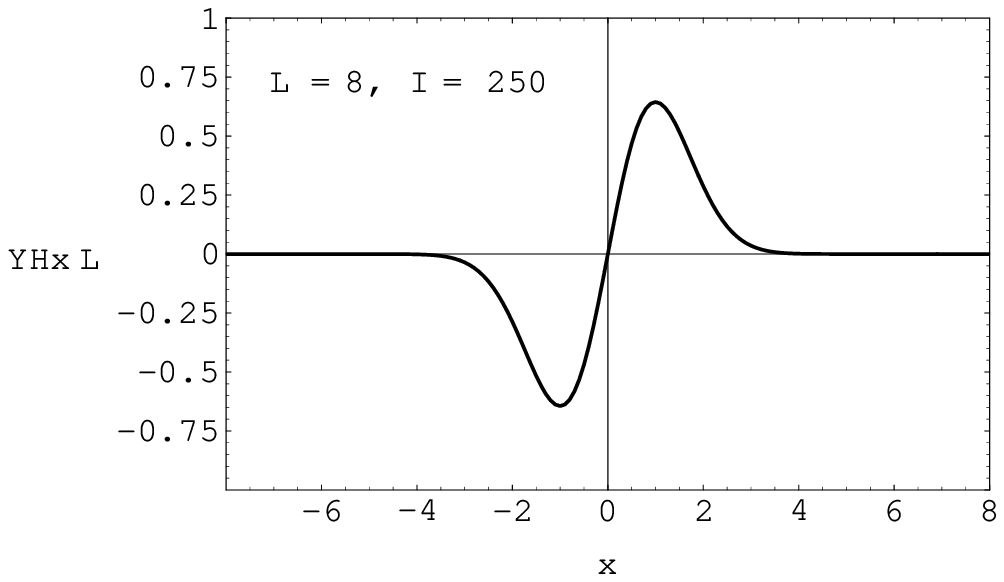}}
\end{minipage}
\caption{{\small The normalized ground (left) and first excited (right) state
wavefunctions for the bounded harmonic oscillator for $L=8$ .}}
\label{figharm}
\end{figure}

In table~\ref{conv} we present the calculated energies for the
ground and the first three excited states for the bounded harmonic oscillator as
compared to the exact results of the unbounded one.
\clearpage
\begin{table}[htbp]
\begin{center}
\begin{tabular}{ccclc}
\hline
\hline
$I$ & $L$ & $N$ & $E_N$ & $E_N^{\mbox{exact}}=(N+\frac{1}{2})$  \\
\hline
$250$&$8$& $0$ & $0.50000000000000000000000000$ & $\frac{1}{2}$\\\\
&  &  $1$ &      $1.500000000000000000000000$ & $\frac{3}{2}$\\\\
%%%%%%%%%%%%%%%%%%%%%%%%%%%%%%%%%%%%%%%%%%%%%%%%
&  & $2$ &       $2.5000000000000000000000$ & $\frac{5}{2}$\\\\
 & & $3$ &       $3.500000000000000000000$  & $\frac{7}{2}$ \\\\
%%%%%%%%%%%%%%%%%%%%%%%%%%%%%%%%%%%%%%%%%%%%%%%%%%%%%%%%%%%%%%%
\hline
\hline
\end{tabular}
\end{center}
\caption{\small The calculated first four energy levels for the
bounded harmonic oscillator compared to the unbounded one.
$(2L)$ is the width of the well and $I$ refers to the number of
the non-vanishing terms in the truncated series of the wavefunction.}
\label{conv}
\end{table}
 The present method, as can be seen from  table~\ref{conv}, reproduces for a large value of
$L$, the exact ones even for a moderate number of non-vanishing terms in
the truncated series of the wavefunction. Moreover, one can get an
accurate description for the wavefunctions shown in
figure~\ref{figharm} which can not be distinguished from the exact
ones when drawn within the same interval $|q|\le L=8$.
%%%%%%%%%%%%%%%%%%%%%%%%%%%%%%%%%%%%%%%%%%%%%%%%%%%%%%%%%%%
%%%%%%%%%%%%%%%%%%%%%%%%%%%%%%%%%%%%%%%%%%%%%%%%%%%%%%%%%%%
%%%%%%%%%%%%%%%%%%%%%%%%%%%%%%%%%%%%%%%%%%%%%%%%
%%%%%%%%%%%%%%%%%%%%%%%%%%%%%%%%%%%%%%%%%%%%%%%%

\begin{figure}[hbt]
\centering
\begin{minipage}[c]{0.5\textwidth}
\epsfxsize=7cm
\centerline{\epsfbox{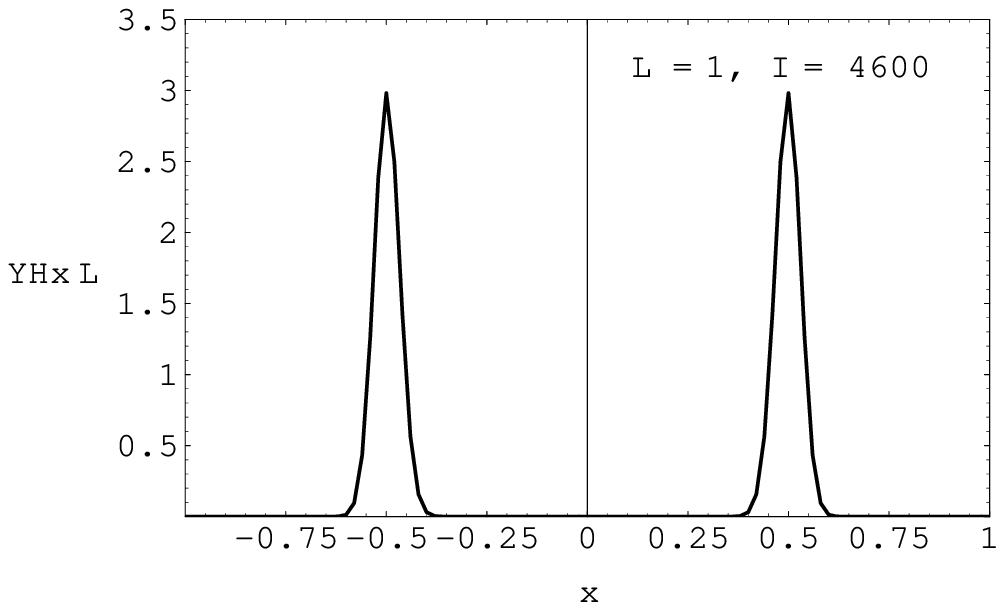}}
\end{minipage}%
\begin{minipage}[c]{0.5\textwidth}
\epsfxsize=7cm
\centerline{\epsfbox{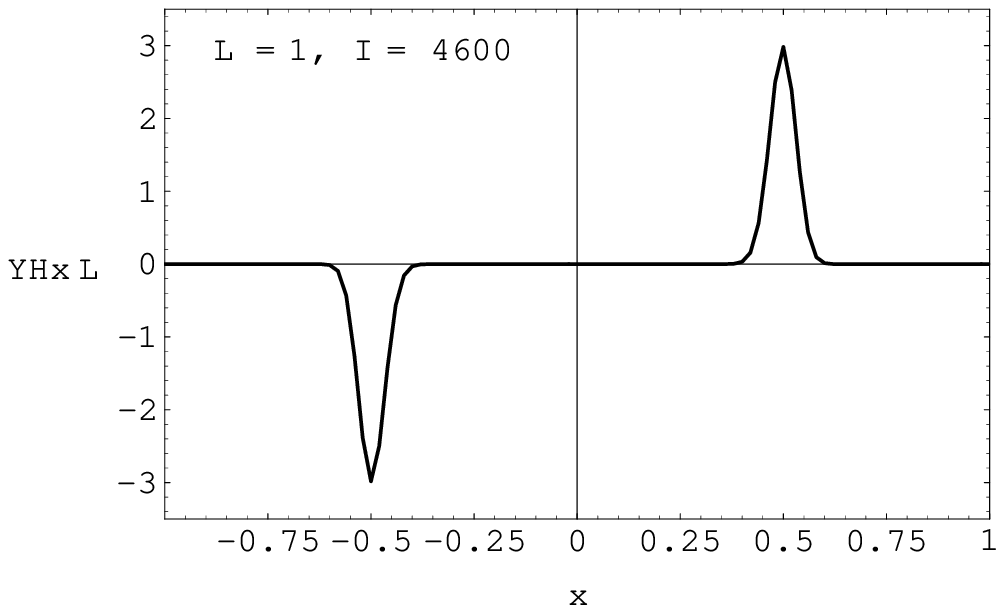}}
\end{minipage}
\caption{{\small The normalized ground (left) and first excited (right) state
wavefunctions for the bounded double-well potential with $g=1/1000$ .}}
\label{figzin}
\end{figure}

Now we apply the above-explained method to the double well
potential in \eq{ham}.
For our convenience, we use the substitution  $q \rightarrow q+{1\over 2}$, so the
potential in \eq{ham} now takes the form
\beq
V(q)={1\over 2} \left(q + {1/2}\right)^2 \left(q-{1/2}\right)^2.
\label{ourv}
\eeq
This form of the potential has now inversion symmetry $(q\rightarrow -q)$ which
is suitable for our calculation. It should be evident that rewriting the
potential in this form doesn't affect the eigenvalues of the Hamiltonian in
\eq{ham}.
As explained above, for this potential we again use the power series expansion of the
wavefunction in the finite range.
The Schr$\ddot{\mbox{o}}$dinger equation, for the potential $V(q)$ in \eq{ourv},
is
\beq
\left[{g\over 2} \frac{d^2}{d\,q^2}+E-{1\over g}V(q)\right]\Psi(q)=0
,\hspace{1cm}
-L < q < L .
 \label{scheq}
 \eeq
In substituting the power series expansion,
\beq
\Psi(q)=\sum_{i} a_i\, q^i ,
\label{wave}
\eeq
in \eq{scheq}, one gets the following
recurrence formula for the expansion coefficients, $a_{i}$:
\beq
a_i=\left({2\over g}\right)\; { {1\over 2 g} \left[a_{i-6}-{1\over 2} a_{i-4}
+{1\over 16} a_{i-2}\right] - E a_{i-2} \over
i(i-1)},\;\; i\neq 0,1\;\;\mbox{and}\;\;
a_i = 0 \;\mbox{when}\; i < 0.
\eeq
For $L=3$, the obtained eigenvalues are
\bea
E_{0,+}(0.001)&=& 0.49899\;54548\;62109\;17168\;91308\;39481\;92163\;68209\;47240\;20809\nnu\\
              & &   \;\;\;66532\;93278\;69722\;01391\;
              \underline{15135\;28505\;38294\;45798\;45759\;95999}\nnu\\
              & &   \;\;\;\underline{06739\;55175\;84722\;67802\;81306\;96906\;01325\;25943\;77289\;94365}\nnu\\
              & &   \;\;\;\underline{88255\;24440\;17437\;12789\;27978\;99793\;98922\;00536\;06978\;04138}\nnu\\
              & &\;\;\;\underline{65255\;73028\;37723\;50241\;67171},
\eea
\bea
E_{0,-} (0.001)&=& 0.49899\;54548\;62109\;17168\;91308\;39481\;92163\;68209\;47240\;20809\nnu\\
              & &   \;\;\;66532\;93278\;69722\;01391\;
              \underline{29839\;92959\;55803\;70812\;27749\;92448}\nnu\\
              & &   \;\;\;\underline{48259\;36743\;64757\;68328\;84835\;35511\;34663\;06309\;82331\;51885}\nnu\\
              & &   \;\;\;\underline{23308\;08622\;84780\;52722\;10103\;67282\;72047\;61340\;01672\;24803}\nnu\\
              & &\;\;\;\underline{65523\;52410\;13798\;16304\;58360}.
\eea
These values agree with the ones obtained from the numerical
calculations based on the Zinn-Justin conjecture. In figure~\ref{figzin} we present
the ground and the first excited state wavefunctions
for the bounded double-well potential for $g=1/1000$, $I=4600$ and $L=1$.
%%%%%%%%%%%%%%%%%%%%%%%%%%%%%%%%%%%%%%%%%%%%%%%%%%%%%%%%%%%
%%%%%%%%%%%%%%%%%%%%%%%%%%%%%%%%%%%%%%%%%%%%%%%%%%%%%%%%%%%
%%%%%%%%%%%%%%%%%%%%%%%%%%%%%%%%%%%%%%%%%%%%%%%%
%%%%%%%%%%%%%%%%%%%%%%%%%%%%%%%%%%%%%%%%%%%%%%%%
\section{Discussion}
It is important to note the following generic remarks. First,
a reason for the capability of the present method is that for a bound
state, the wavefunction is spatially localized which means that the probability density
$(|\Psi|^2)$ has appreciable values in a finite region of space behind which the
probability density tends rapidly to zero. Thus, to a good approximation, it is,
therefore, reasonable to consider the corresponding problem in a finite interval,
with a suitable width, bounded by two infinite walls. The criteria for a suitable
value of $L$ can be quantitatively  given by the condition $E << V(L)$.
Second, from the WKB approximation, it can be made plausible that
the zeros of $\Psi(E,L)$ provide upper bounds for the
energy eigenvalues while the zeros of the derivative--with respect to $q$--
$\Psi'(E',L)$ provide the lower ones; the same finding can be proved in a rigorous way
as shown in \cite{tas}. Thus, by matching the digits of the two
zeros, one can get an accurate energy eigenvalue up to the number of
coincident digits. As an example for the ground state of the bounded harmonic
oscillator, with $L=8$ and $I=250$, one gets
\bea
E&=&0.5000000000000000000000000014362707054755765903756598\nnu \\
  & & 26757972824824621785332078167891514939744867648,\nnu\\
E'&=&0.4999999999999999999999999985405543573278682092744652\nnu\\
& &    58622103903146216005437303539479001558808137418.\nnu\\
\label{eep}
\eea
The corresponding wavefunctions and their slopes are
\bea
\Psi(E,8) = 4.8\times 10^{-49}&,& \Psi'(E',8)=8.1\times
10^{-48},\nnu \\
\Psi(E',8) = 2.6\times 10^{-14}&,& \Psi'(E,8)=-2\times
10^{-13}.\nnu \\
\label{zero}
\eea
After matching the digits of the two  numbers in \eq{eep}, one gets
the ground-state energy accurate up to $25$ digits as shown in
table~\ref{conv}. The remaining eigenvalues are obtained by the
same procedure.
However, one should pay attention that this accuracy
is expected to be less than the accuracy of the bisection
method. In this method, the accuracy estimation is $\varepsilon = (c-a)/2^n$
where $n$, here, is the number of iteration, and $c$ and $a$ are two points enclosing only one root.
In our case, we have taken for the ground state $n=200$, $c={6\over 10}$ and $a={4/10}$ ,
giving  $\varepsilon = 1.2\times 10^{-61}$. Finally
according to the WKB approximation, the wavefunction behaves for large $q$ in
the inaccessible region as
\beq
\Psi_{WKB}(q)\;\propto {1\over (V(q)-E)^{1\over 4}}\,
\exp{\left(-\int^q_{q_t}\sqrt{(V(q')-E)}\,dq'\right)},
\label{wkb0}
\eeq
where $q_t$ is a turning point just left to the inaccessible region.
The value of $\Psi_{WKB}(q=8)\;\mbox{is}\; 6.5\times 10^{-14}$, while for the
truncated series solution of \eq{sol0} it has the value $4.8\times 10^{-49}$ as given
in \eq{zero}.
The reason for this huge difference is that the series solution is valid
and convergent as long as $q$ is finite \cite{codd}. In addition to this,
the energy eigenvalues as extracted from the
zeros of $\Psi(E,L)$ (for suitable $L$) result in a delicate
cancellation between terms of opposite signs in the power series solution.

One may suspect that  using  a series solution in the form
\beq
\Psi(x)=\exp{(-b\, x^2)}\sum_{j} a_j \, x^{j},
\label{serexp}
\eeq
may help improving the rate of convergence for the obtained
eigenvalues. In contrast, one needs more terms in the series expansion to
achieve the same level of accuracy obtained by the series solution of the form
given in \eq{sol0}. The reason behind this stems from the fact that any
finite truncation for the  series in the form given in \eq{serexp} always decays,
due to the exponential factor, as $q$ becomes large  making the determination of
the energy eigenvalues less reliable, especially when the parameter $b$ is large.
As an example, when $b={1\over 2}$ we can achieve the same accuracy
reported in table~\ref{conv} with the same number of
non-vanishing terms in the truncated series, while for $b=8$ we
need 600 non-vanishing terms to achieve the same accuracy. Thus,
the best thing which can be done is to work with the parameter $b$ having
zero value. However, it should be kept in mind that both series in \eq{sol0} and
\eq{serexp} are equivalent  but only in the infinite sum limit.

We also study the effect of the parameter $b$ in the case of the double-well
potential given by
\beq
V(x)=-10\, x^2 + x^4,\;\;\;\;\;\; \left(\mbox{in units}\;\;\;
\hbar = 1, m= \frac{1}{2}\right).
\label{dwpot}
\eeq
As an example, when we work with the precision $100$ digits, then we find for
$b=0,\; I= 750\; \mbox{and}\;\;L=8$,
that the ground state energy has the value (accurate up to $69$ digits)
\bea
E_0&=& -20.63357\;67029\;47799\;14995\;85548\;37431\;\nnu\\
& & 50876\;53159\;46057\;73551\;39057\;10311\;42892\;92.
\eea
To achieve the same accurate energy determination for $b=10$, we
find that it is possible to use $500$ terms which is not considerably
less than the case of $b=0$. However, this comes with the high cost of working with
precision $300$ digits. Working with such a high precision renders the calculation slow. At
intermediate values of $b$ like $2,3$ and $4,$ we can use less
terms but with high precision as shown in table~\ref{precision}.
According to our numerical investigations for the case of
the double well, in the finite range, the choice $b=0$ is the best
compromise between
the number of terms used and the degree of precision  to get a more
efficient calculation.
%%%%%%%%%%%%%%%%%%%%%%%%%%%%%%%%%%%%%%%%%%%%%
\begin{table}[hbtp]
\begin{center}
\begin{tabular}{ccccccccc}
\hline
\hline
$b$ & $0$ & ${1\over 2}$ & $1$ & $2$ & $3$ & $4$ & $5$ & $10$   \\
\hline
$I$&$750$& $750$ & $750$ & $500$ & $500$ & $500$ & $500$ & $500$ \\
\hline
$\mbox{Precision}$&$100$&$100$&$150$&$150$&$200$& $200$& $200$ &
$300$\\
\hline
\hline
\end{tabular}
\end{center}
\caption{\small Precision versus $I$ and the parameter $b$ }
\label{precision}
\end{table}

It is important to point out that in dealing with low accuracy results (like nine digits),
one cannot decide which is better, to work with or without the parameter $b$.
Furthermore, employing the method in a non-efficient way may
lead to wrong conclusions as in \cite{kil}, where it is emphasized
that setting a non-vanishing value for the parameter $b$ greatly reduces
the number of terms used. To clarify these points, we obtain
for the potential given by \eq{dwpot}
the four first energy levels $(E_0=-20.6335767,\;E_1=-20.6334568,\;
E_3=-12.3795437,\;E_4=-12.3756738)$ accurate up  to $10$ digits as presented in
\cite{kil}; our  results (using $L=4.2$) are summarized in table~\ref{num}.
\begin{table}[hbtp]
\begin{center}
\begin{tabular}{ccccccccc}
\hline
\hline
$b$ & $0$ & ${1\over 2}$ & $1$ & $2$ & $3$ & $4$ & $5$ & $10$   \\
\hline
$I$&$125$& $100$ & $90$ & $90$ & $90$ & $90$ & $90$ & $200$ \\
\hline
\hline
\end{tabular}
\end{center}
\caption{\small The parameter $b$ versus I (number of non-vanishing terms)}
\label{num}
\end{table}
It is evident from table~\ref{num} that one can not say it is a
big advantage to use $90$ terms (for $b=2$) rather than $125$ terms
(for $b=0$). However, numerical studies clearly indicate that the
situation becomes worse when $b$ increases (for $b=10$ we need $200$
terms). Another clear example is the pure quartic potential $(V(x)=x^4)$
for which we get, for
$b=0,\; L=3.5,\; \mbox{and}\; I=75$, low-energy eigenvalues (the first
five) determined accurately up to nine digits while obtaining the same
results for the choice $b=3$ and $I=50$. Furthermore, the tenth eigenvalue
is determined accurately up to $9$ digits, for $L=3.9$, using $I=75$
for $b=3$, while $I=125$ for $b=0$. These findings are in
contradiction with what has been claimed in~\cite{kil}, where
it was mentioned that one should use about $2000$ terms in the
power series to determine the energy for the choice $b=0$. Similar
findings occur for the potential $V(x)=x^2 + x^8$. In such a
situation for $L=2.5$, we can use $125$ terms in the power series
solution for $b=0$ and $75$ terms for $b=5$, while
getting the same accurate results up to nine digits.

The problem in the calculations found in \cite{sec,kil}
comes from evaluating every term in the power series to a certain precision, and
then summing the series which leads to an error accumulation,
resulting in low-accuracy results despite using a large number of terms.
In our approach, we sum all terms in the power series  exactly, and then only
in determining the roots (energy) from the condition $\Psi_{I}(E,L)=0$, do we resort
to numerical calculation with a certain precision.
Although the ability of the computer algebra system to deal with exact numbers
was available from the early $1980\mbox{s}$, it has not been used since
then in such calculations.
%%%%%%%%%%%%%%%%%%%%%%%%%%%%%%%%%%%%%%%%%%%%%%%%%%%%%%%%%%%%%%%
\section{Conclusion}
In this paper we have presented an independent simple method leading
to eigenvalues which agree well with the recently obtained numerical results
based on the Zinn-Justin conjecture for the symmetric double-well potential.
We have also included results with more significant
digits than reported. It has been applied to some other potentials
to illustrate its capability, and its precision has been compared
with other calculations based on introducing an exponentially
decaying factor $(e^{-\,b\,x^2})$. Several subtle points related
to its precision have also been discussed and clarified.
The method we opted also enables us to get an accurate numerical
determination of the corresponding wavefunctions.
\begin{flushleft}
{\bf Acknowledgement}
\end{flushleft}
This work was supported by Research Center at College of Science,
King Saud University under project number Phys$/1423/02$.
%\newpage
%------------------------- REFERENCES ------------------------------
\renewcommand{\baselinestretch}{1}

\end{document}